\documentclass[fleqn,usenatbib]{mnras}

\usepackage{newtxtext,newtxmath}

\usepackage[T1]{fontenc}
\usepackage{ae,aecompl}

\usepackage{graphicx}	
\usepackage{amsmath}	
\usepackage{amssymb}	

\title[Circular polarization in IGR J17014$-$4306]{Discovery of spin modulated circular polarization from IGR J17014$-$4306, the remnant of Nova Scorpii 1437 A.D.}

\author[Stephen B. Potter \& David A. H. Buckley]{
Stephen B. Potter$^{1}$\thanks{E-mail: sbp@saao.ac.za} \&
David A. H. Buckley,$^{1}$
\\
$^{1}$South African Astronomical Observatory, PO Box 9, Observatory, 7935, Cape Town, South Africa\\
}

\date{Accepted XXX. Received YYY; in original form ZZZ}

\pubyear{2015}

\hypersetup{draft}

\begin{document}
\label{firstpage}
\pagerange{\pageref{firstpage}--\pageref{lastpage}}
\maketitle

\begin{abstract}
Polarimetry of IGR J1401$-$4306, a long period (12.7 hours), eclipsing intermediate polar and remnant of Nova Scorpii 1437 A.D., reveals periodic variations of optical circular polarization, confirming the system as the longest period eclipsing intermediate polar known. This makes it an interesting system from an evolutionary perspective. The circular polarization is interpreted as optical cyclotron emission from an accreting magnetic white dwarf primary. Based on the polarimetry, we propose that it is a disc-fed intermediate polar. The detection of predominantly negative circular polarization is  consistent with only one of the magnetic poles dominating the polarized emission, while the other is mostly obscured by the accretion disc.
\end{abstract}



\section{Introduction}

Cataclysmic variables (CVs) are interacting binaries in which gas from a Roche lobe filling, late-type, main-sequence star (the secondary) is being transferred to a white dwarf (the primary). In general the material from the secondary spirals down to the white dwarf through an accretion disc. In the intermediate polar (IP) class of magnetic CVs the magnetic field of the white dwarf is sufficiently strong to disrupt and truncate the inner part of the accretion disc. From this point on accretion proceeds via accretion curtains on to regions near the magnetic poles of the white dwarf (see e.g. \cite{hellier1999} and references therein). The gas becomes shocked to high temperatures, emitting X-rays and sometimes optically polarized cyclotron radiation in the optical/infrared region.

In classical novae, prior to eruption, material is almost continuously being deposited on to the surface of the white dwarf at a very high rate. At some stage a critical density of accreted material is reached, which results in a thermonuclear runaway in the hydrogen envelope of the primary. These events are observed as nova explosions, with material being expelled from the binary system \citep{Starrfield1972, Starrfield1976, Prialnik1978}. The brightness of these novae can increase from 6 to 19 magnitudes. Comprehensive reviews on classical novae and CVs can be found in \citep{bode1989} and \citep{Warner1995}, respectively. 

Dwarf novae, on the other hand, have much lower mass transfer rates and exhibit periodic dwarf nova outbursts when accretion-disk instabilities cause the discs to become hot, optically thick and brighter. They are observed to have multiple eruptions on timescales of weeks to years, increasing in brightness from 2 to 7 magnitudes.

CVs continue to be discovered as novae  eruptions (classical and dwarf) or by X-ray satellites (e.g. {\it ROSAT, ASCA, RXTE, BeppoSAX, INTEGRAL}). Magnetic CVs, in particular, were discovered as a result of their X-ray emissions. 

The hard X-ray source, IGR J17014$-$4306, discovered by the {\it INTEGRAL} satellite \citep{krivonos2012}, was suggested to be a magnetic CV on the basis of followup optical spectroscopy \citep{Masetti2013}. Further X-ray and optical photometry by \cite{Bernardini2017} and optical photometry by \cite{Shara2017} showed strong evidence that the system was an intermediate polar with a $\sim$1859 s spin period.

\cite{Shara2017} reported the recovery of the CV remnant of the classical nova of 11 March 1437 A.D. (recorded by Korean royal astronomers) as IGR J17014$-$4306, independently confirming its age by proper motion-dating. They also show that, in the 20th century, it exhibited three dwarf nova type eruptions in 1934, 1935 and 1942. The CV shows ellipsoidal variability and is deeply eclipsing which enabled the measurement of its orbital period $P_{orb}=0.5340263 \pm 5E-7$ days, and detailed characterization of its stellar components. From their optical photometry and Chandra X-rays they detect (using Discrete Fourier analysis) a period of 1859.112 $\pm$ 0.069 seconds, which they also attribute to the spin period of the white dwarf, which is more than 5$\sigma$ from the period derived by \cite{Bernardini2017}, namely 1858.67 $\pm$ 0.02 s derived from AAVSO optical V-band observations only (using a two composite sinusoidal model for the fundamental
and its first harmonic.) 


\section{Observations}

\begin{table}
	\centering
	\caption{Table of observations. All observations were made with the HIgh-speed-Photo-Polarimeter (HIPPO; \citep{potter2010}) on the 1.9m telescope of the South African Astronomical Observatory}
	\label{tab:example_table}
	\begin{tabular}{lccr} 
		\hline
		Date & No.Hours & Filter(s) & Eclipse\\
		\hline
		25 June 2017 & 8.1 & clear & No\\
		30 June 2017 & 0.85 & BG39, OG57 & Yes\\
		30 June 2017 & 2.2 & OG57 & No\\
		13 July 2017 & 3.84 & OG57 & No\\
        14 July 2017 & 4.32 & OG57 & Yes \\
        16 July 2017 & 0.54 & OG57 & No\\
        17 July 2017 & 5.1 & clear & No\\
        18 July 2017 & 4.7 & OG57 & No\\
        22 July 2017 & 6.5 & clear & Yes\\
		\hline		
	\end{tabular}
\end{table}

IGR J17014$-$4306/Nova Sco 1437 A.D. was observed on 9 nights of June and July 2017 (see Table 1). All observations were made with the HI-speed Photo-POlarimeter (HIPPO; \cite{potter2010}) on the 1.9-m telescope of the South African Astronomical Observatory. The HIPPO was operated in its simultaneous linear and circular polarimetry and photometry mode (all-Stokes). Observations were either clear filtered (3500-9000 \AA) defined by the response of the two RCA31034A GaAs photomultiplier tubes or through broad-band blue (BG39) and red (OG570) filters.

Several polarized and non-polarized standard stars \citep{Hsu1982, bastien1988} were observed in order to calculate the waveplate position angle offsets, instrumental polarization and efficiency factors. Photometric calibrations were not carried out; photometry is given as background sky subtracted total counts. Background sky photometry and polarization measurements were taken at frequent intervals during the observations.

All of our observations were synchronized to GPS to better than a millisecond. Given the high-speed nature of HIPPO, its timing accuracy has been verified through feeding GPS pulsed LED light fed through the instrument.  

We corrected all times for the light travel time to the barycentre of the Solar system, converted to the barycentric dynamical time (TDB) system as Barycentric Julian Date (BJD; see \cite{Eastman2010}, for achieving accurate absolute times and time standards). By doing this we have removed any timing systematics, particularly due to the unpredictable accumulation of leap seconds with UTC, and effects due to the influence of primarily Jupiter and Saturn when heliocentric corrections only are applied. Data reduction then proceeded as outlined in \cite{potter2010}.

\begin{figure*}

    \centering
	\includegraphics[width=\textwidth]{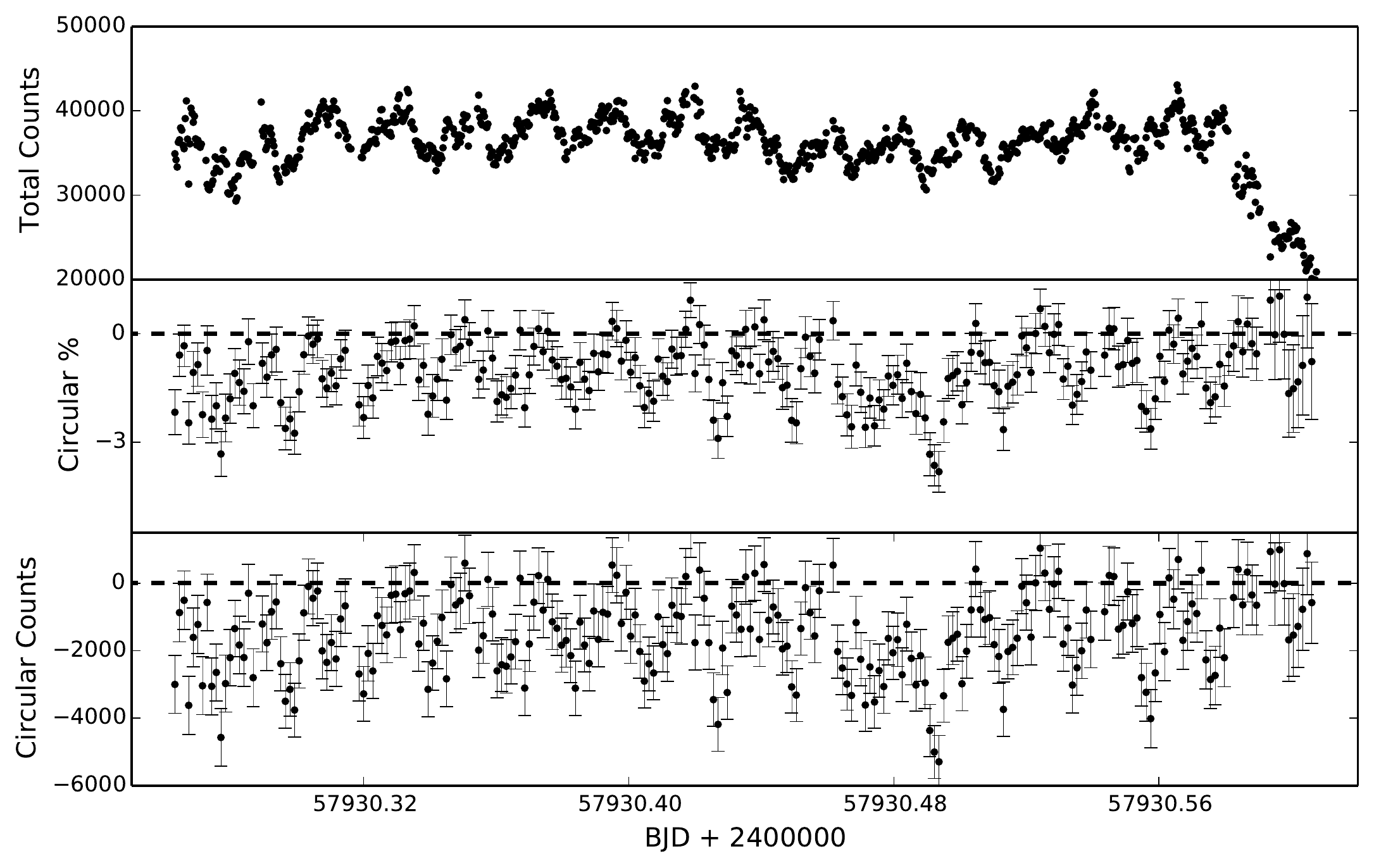}
    \caption{Photo-polarimetry from 25 June 2017. Top to bottom panels correspond to the photometry (30s bins), percentage of circular polarization (120s bins) and circular polarized counts (120s bins). Worsening seeing is the cause of the photometric drop off towards the end.}
    \label{fig:example_figure}
\end{figure*}

\section{Results}

\begin{figure}
	\includegraphics[width=\columnwidth]{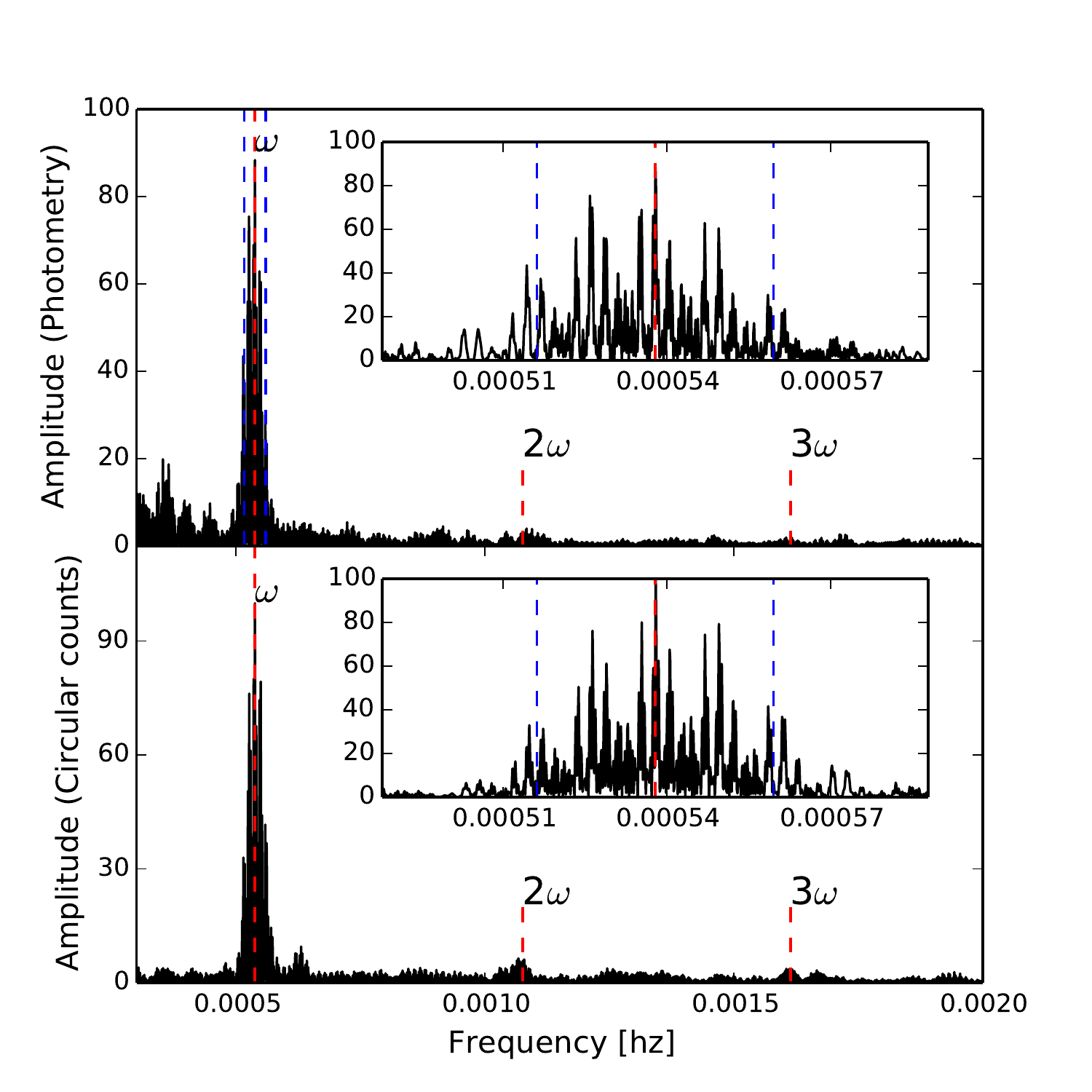}
    \caption{Fourier analysis of the combined data set: Top and bottom panels are the amplitude spectra of the photometry and circular polarized counts respectively. $\omega$ is the spin frequency, also indicated by the red, vertical, dashed line. Blue, vertical, dashed line indicate where the beat frequency would be expected. Inserts show expanded views around the spin frequency. }
    \label{fig:example_figure}
\end{figure}

\begin{figure}
	\includegraphics[width=\columnwidth]{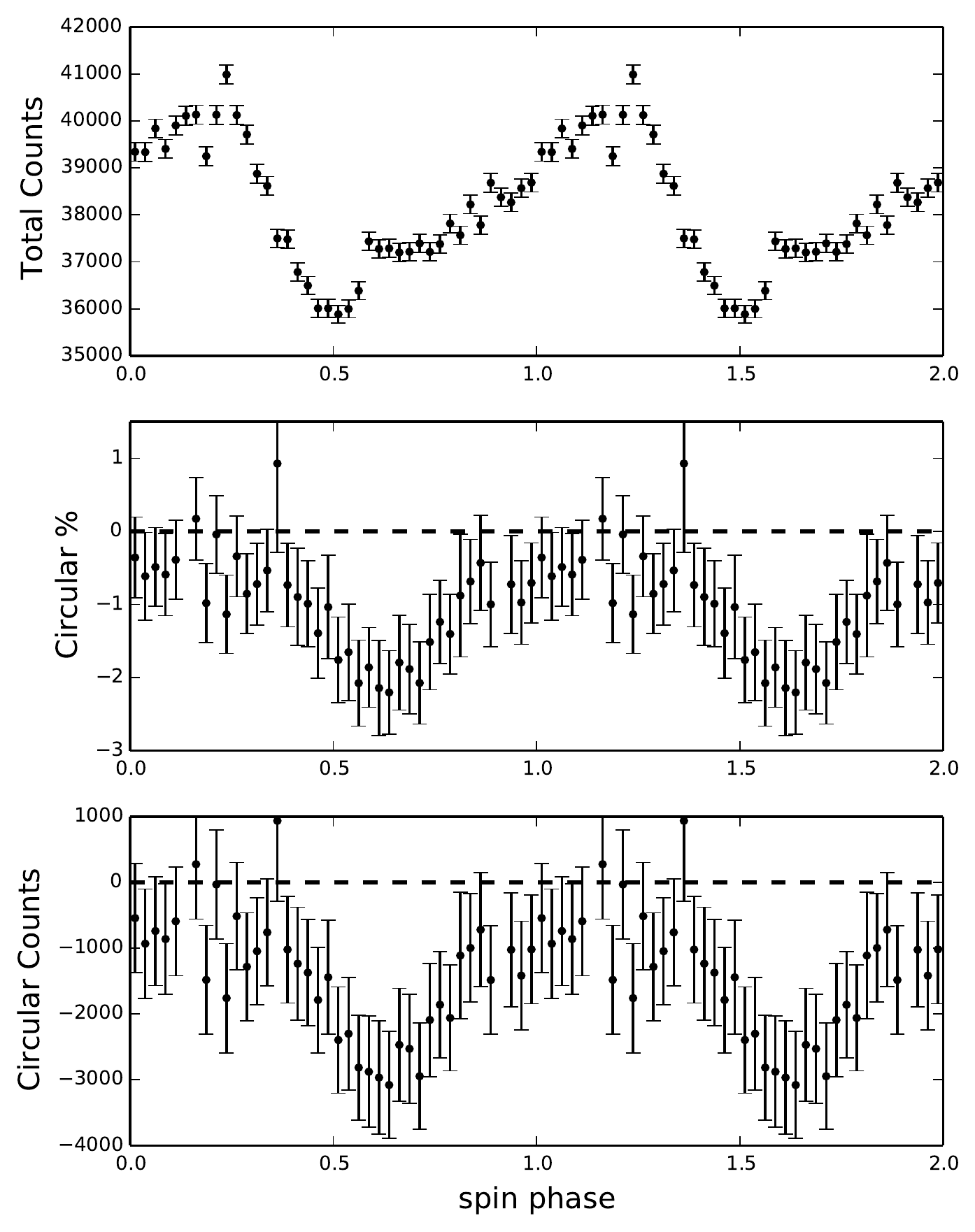}
    \caption{Spin folded photometry and circular polarimetry of the June 25th observations. Bin sizes correspond to  $\sim$ 46s (0.025 spin phase bins) for  the photometry and polarimetry.}
    \label{fig:example_figure}
\end{figure}

Fig. 1 shows our longest observation, specifically the clear-filter photopolarimetric observations taken on the night of 2017 June 25/26 for a total of $\sim$ 8.1 hours.  Conditions were  photometric except for the final  $\sim$ 30 minutes  where some light loss occurred during worsening seeing. Using the orbital ephemeris of \cite{Shara2017} these observations correspond to an orbital phase range between $\sim$0.04-0.69($\pm$0.0006). 

The photometry has not been flux calibrated, however its relative brightness compared to the surrounding field stars, as observed in the telescope acquisition image, appears consistent with the optical image presented in \cite{Masetti2013}, giving an R magnitude of $\sim$15-16. 

The 1859s white dwarf spin modulation can clearly be picked out in the clear-filter photometry (binned to 30s, Fig.1 top panel).  The clear-filtered circular polarimetry (binned to 120s) shows variability mostly confined between 0 and $-$4 per cent. There are some excursions to positive circular polarization although the S/N precludes a definite detection. Fig. 1 bottom panel shows the circularly polarized counts. The circular polarization also appears spin pulsed with maximum percentage (and counts) occurring near the minimum of the photometric pulses.

All of the photometry and polarimetry as listed in Table 1 were subjected to Fourier analysis. Fig. 2 presents the amplitude spectra. The top and bottom panels display the photometry and  circular polarimetry respectively. 

The amplitude spectra of the clear filtered photometry is dominated by a singular peak coincident with the spin frequency reported by \cite{Bernardini2017} and \cite{Shara2017}. The expanded view, centered on the spin frequency, shows multiple peaks as a result of aliasing. None of these peaks are consistent with the expected location of the beat frequency (spin frequency - orbital frequency) corresponding to a beat period of $1937.1(1.3)$s. Harmonics are also present at 2 and 3 times the spin frequency (shown as vertical dashed lines) but at a very low level.

The corresponding circular polarimetry also has a significant peak coincident with the spin period, thus confirming the intermediate polar status of IGR J17014$-$4306/Nova Sco 1437 A.D. through the discovery of spin modulated circular polarization. Similar to the photometry, small peaks are also present at 2 and 3 times the spin frequency (shown as vertical dashed lines). The expanded view, centered on the spin frequency, shows multiple peaks as a result of aliasing. None of these peaks are consistent with the expected location of the beat period. Both the photometry and circular polarimetry peaks have maximum values centered on $1859.1(1.3)$s. The error represents the FWHM of the peak which precludes distinguishing between the periods of \cite{Bernardini2017} and \cite{Shara2017}.

Linear polarization is seen at a level of $\sim$1.0 per cent (not shown); however, the S/N is not sufficient to claim a firm detection and is probably due to interstellar polarization. In addition, no significant periods were detected in the Fourier analysis of the linear polarimetry. 

The spin modulation is confirmed in Fig. 3, where the clear filtered, 25 June, photometry and polarimetry have been spin-phase-fold-binned on the spin ephemeris of \cite{Shara2017}.  The accumulated  error of the spin ephemeris amounts to  $\sim$ 0.0125 days which is more than half a spin period.  Therefore it is not unexpected to find that the maximum flux of the photometry is not located at phase 0, as defined by the spin ephemeris. The accumulated error also precludes the refinement of the spin period due to ambiguous cycle counts.

\cite{Shara2017} noted two pulses, with different amplitudes, per spin period in their photometric observations, which would produce significant power at the first harmonic. They further mentioned that the pulse profiles changed slightly from night to night. The photometric pulse profile, presented in Fig.  3, displays a more single saw-tooth shape, rather than a double-peaked profile. There is some indication of a changing pulse profile throughout our datasets, however we do not have sufficient observations to derive further conclusions.

The circular polarization  appears to be negatively polarized throughout the full spin cycle with  a single pulse of $\sim$ half a spin period.  The peak of the polarized pulse occurs $~\sim$ 0.1 of a spin phase after the  minimum of the photometric pulse.

Fig. 4 shows the 14 (OG570) and 22 (CLEAR) July 2017 observations of the eclipses of IGR J17014$-$4306/Nova Sco 1437 A.D. The predicted times of eclipse according to \cite{Shara2017} are indicated by vertical dashed lines while the span of the eclipses is indicated by the shaded grey regions. The eclipse morphology looks somewhat different for the two observations, where the egress seems more step-like for the 22 July. We believe this is indicative of cycle-to-cycle changes rather than a wavelength effect, since an eclipse observation done on 30 June (not show), taken with the OG570 and CLEAR filter simultaneously, showed exactly the same light curve morphology, both with similarly sharp egresses. 

During the period of eclipse the percentage of polarization and the amount of circular polarized counts is reduced, demonstrating that the negatively polarized cyclotron emission region(s) are being occulted during the whole eclipse. There is some indication that the net circular polarization is positive during the clear filter eclipse, possibly evidence that the second accreting pole is not eclipsed. However the S/N is too low for a firm conclusion and further observations during eclipse would be required to investigate this. 

Outside of eclipse the circular polarized pulses can clearly be seen, similar to those seen in Fig 1. In addition there appears to be variability of the order of $\sim$0.01-0.02 days in the average polarization. For example, the polarization after the OG57 filter eclipse (Fig. 4) is generally more negative than before eclipse and similarly for a region of time before the CLEAR filter eclipse (Fig 4). 

\begin{figure*}
\includegraphics[width=\textwidth]{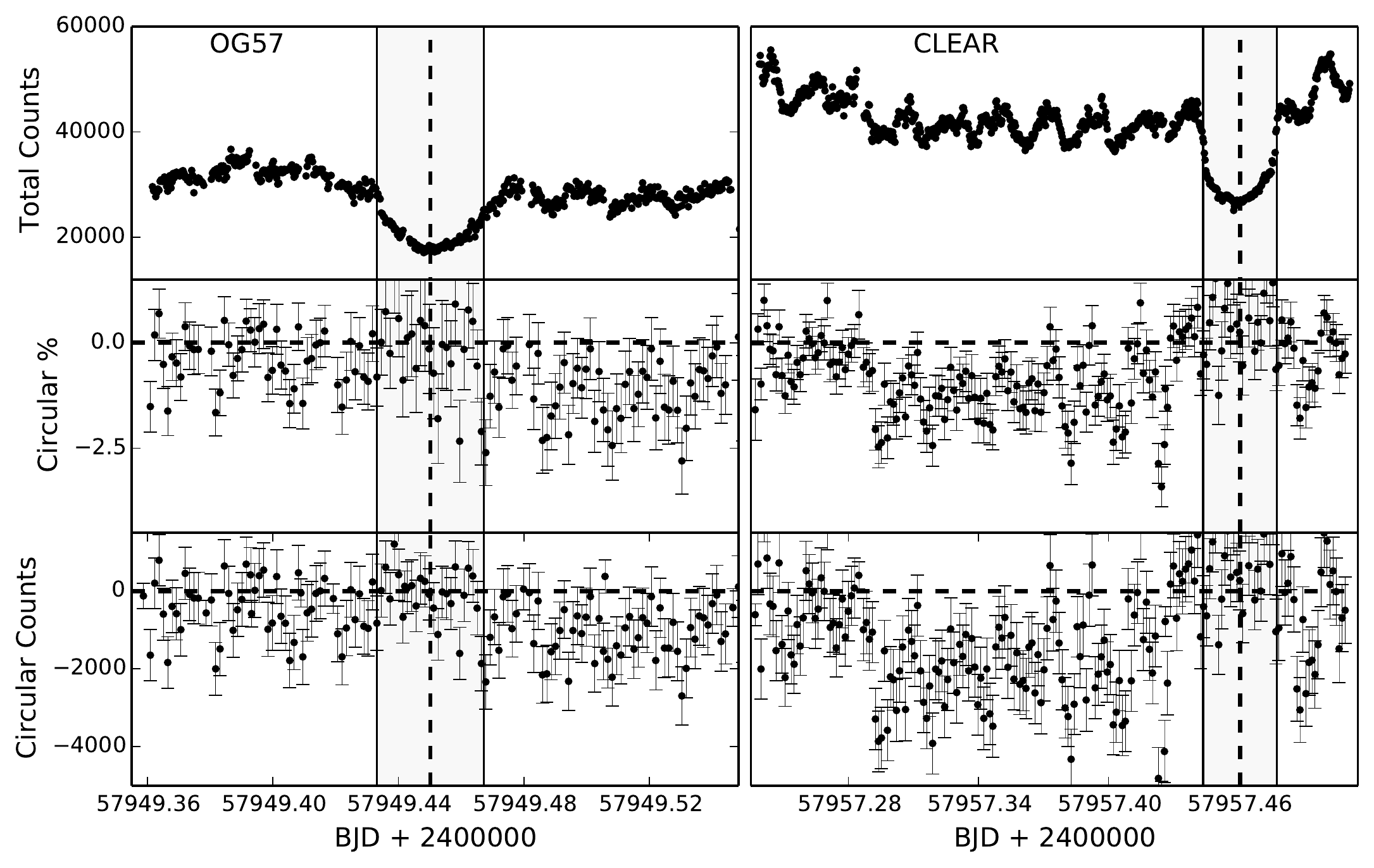}
    \caption{Photo-polarimetry around times of eclipse. Left and right columns show OG57 and clear filtered observations from 14 and 22 July 2017 respectively. Top to bottom panels show the photometry (30s bins), circular percentage and circular counts (120s bins) respectively. Vertical dashed line indicates time of mid-eclipse according to ephemeris of \citep{Shara2017}. }
    \label{fig:example_figure}
\end{figure*}

\section{Discussion}

There are only two other IPs with longer orbital periods than IGR J17014$-$4306/Nova Sco 1437 A.D., namely GK Persei (47.9 hr), see e.g. \cite{crampton1986, bianchini2003} and 1RXS J173021.5-055933 (15.42 hr), see \cite{butters2009}. 

GK Persei is also a classical novae (Nova Persei 1901)  surrounded by the nova remnant; the Fireworks nebula. It was classified as an intermediate polar after the discovery of 351s pulsed X-ray emission, see \citep{mauche2004} and references therein. 

1RXS J173021.5-055933 (IGR J17303-0601, V2731 Oph) was selected from the {\it ROSAT} Bright Source Catalogue \citep{voges1999} by \cite{gaensicke2005} and identified as an Intermediate polar with an orbital period of 15.42 hr and  WD spin period of 128.0s.  UBVRI polarimetric observations were carried out by \cite{butters2009} who reported possible spin modulated circular polarization in the B-filter but not in the other filters. 

IGR J17014$-$4306/Nova Sco 1437 A.D. is now the longest period eclipsing intermediate polar known, which makes it an interesting system from an evolutionary perspective (see Fig. 5).

\subsection{The accretion geometry}

With an orbital period of $>$3h and a $P_{spin}/P_{orb}$ = 0.04, IGR J17014$-$4306/Nova Sco 1437 A.D. is a 'regular' disc-fed IP according to the prescription by \cite{norton2004} and \cite{norton2008}, albeit with an extreme  orbital period of  12.8 hours.

\cite{ferrario1999} have shown that, for stream-fed accretion, significant power would be expected in the optical at the beat frequency ($\omega - \Omega$). In contrast, for disc accretion, the dominant power in the continuum and line fluxes is always at the spin frequency $\omega$. We have found significant power at $\omega$ only, consistent with the 'regular' disc-fed IP classification.  This is further supported by the broad (FW$\sim$2000-2700 km/s) emission of HI Balmer, HeII and HeI lines reported by \cite{Shara2017}.

IGR J17014$-$4306/Nova Sco 1437 A.D. displays negative circular polarization only, which would ordinarily be attributed to either single pole accretion or a viewing geometry that somehow hides one of the magnetically accreting poles. Single pole accretion is difficult to reconcile given the disc-fed classification, which would be expected to produce accretion curtains onto the magnetic poles of both hemispheres.  It is also difficult for the 'lower' pole to be mostly self occulted by the WD given the high inclination. One possibility  is that any emission from the lower pole is obscured by the accretion disc. This scenario has been suggested by \cite{evans2007} in order to partly explain  why some IPs show  soft X-ray emission and others do not.  They reason that the viewing geometry to the so called soft IPs is such that the accreting poles are viewed directly and therefore the soft emission  is not obscured. Hard IPs however can have one or both poles either obscured by an accretion disc or by accretion curtains, which would absorb the blackbody component of the soft emission. Similarly this could also explain why only a handful of IPs show polarized emission, namely that for most inclinations, both poles are relatively un-obscured and will result in a net polarization close to zero. 

The soft IPs are indicated in Fig. 5 with star symbols and the remaining IPs with open circles. The polarized IPs are additionally annotated with a 'P'. Of the soft IPs 63$\%$ (7/11) are polarized, whereas only 8$\%$ (3/37) of the remaining IPs show polarization. This tends to favor the scenario that polarization and soft X-ray emission can occur when the geometry is such that there is an unobstructed view of at least one of the accreting poles during the spin cycle of the white dwarf \citep{evans2007}. IGR J17014$-$4306/Nova Sco 1437 A.D. has a soft blackbody X-ray component \citep{Bernardini2017} and is therefore a soft IP.

\begin{figure}
	\includegraphics[width=\columnwidth]{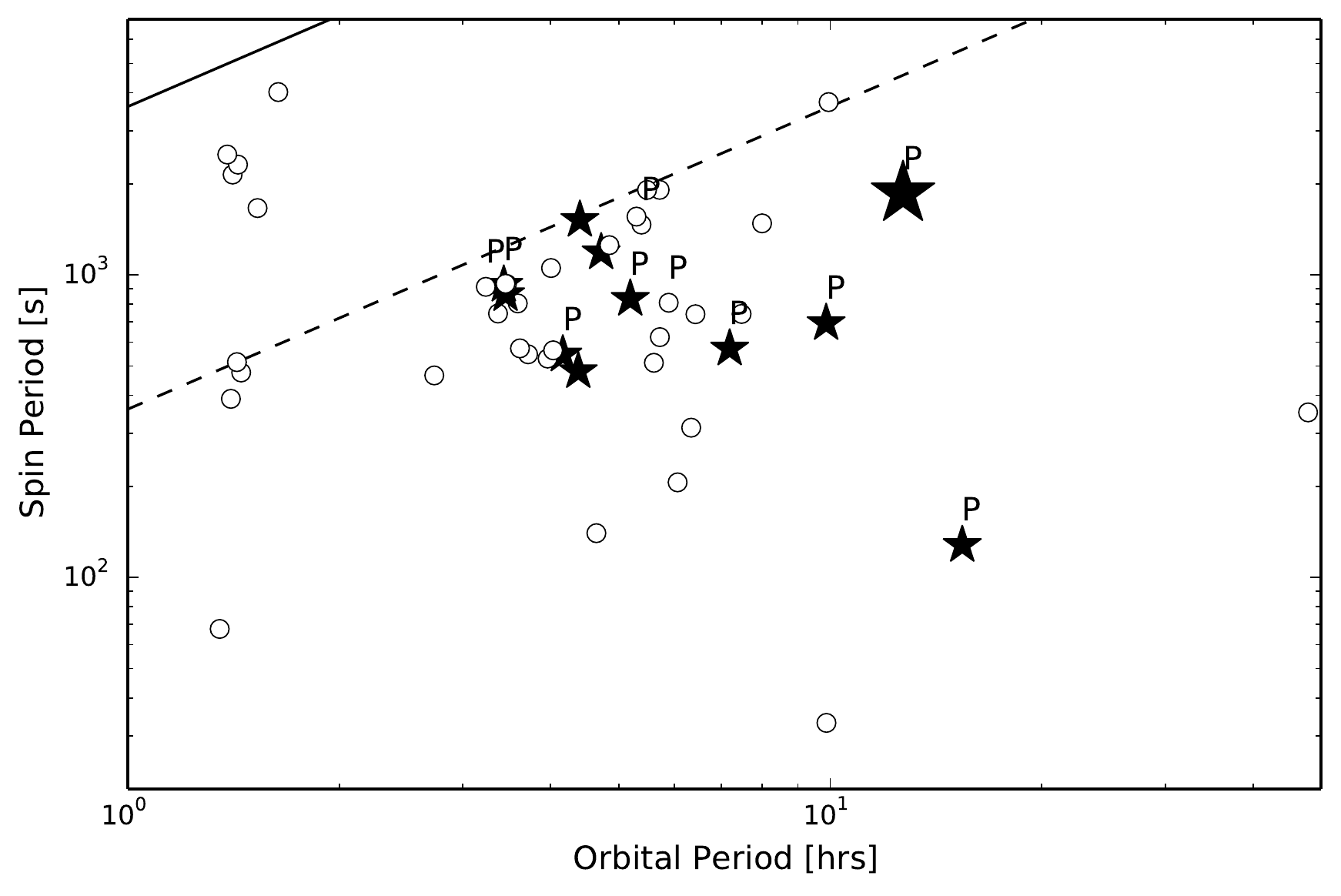}
    \caption{Orbital verses spin period of the so called ironclad intermediate polars. List taken from https://asd.gsfc.nasa.gov/Koji.Mukai/iphome/iphome.html. Large star symbol corresponds to IGR J17014$-$4306/Nova Sco 1437 A.D. Smaller star symbols correspond to the other known soft IPs. Remaining IPs indicated with open circles. Polarized IPs are additionally annotated with a 'P'. Diagonal lines are $P_{orb}=P_{spin}$ (solid) and $P_{orb}=10xP_{spin}$ (dashed).}
    \label{fig:example_figure}
\end{figure}

\subsection{Outbursts}

\cite{Shara2017} have located the CV associated with the remnant of Nova Sco 1437 A.D. from Harvard DASCH (Digital Access to a Sky Century \@ Harvard) photographic plates and identified possible dwarf nova outbursts in the 1930s and 1940s.  Our circular polarimetry identifies Nova Sco 1437 A.D. as an IP rather than a dwarf nova. Nevertheless IPs have been reported to undergo 'outbursts'; e.g. EX Hya \citep{mhlahlo2007,hellier2000} XY Ari \citep{hellier1997}, GK Per \citep{sabbadin1983, watson1985, morales1996}, TV Col \citep{szkody1984,hellier1993}, V1223 Sqr \citep{amerongen1989} and YY Dra \citep{szkody2002}. XY Ari, GK Per and YY Dra are said to have their outbursts caused by an instability in the disc \citep{hellier1997} while other IPs, e.g. TV Col, V1223 Sqr and EX Hya, seem to pose problems for the disc instability model. It has been suggested that the outbursts in these three stars are a result of increased mass transfer from the secondary \citep{hellier1993, hellier1989,hellier2000} due to enhanced mass transfer as a result of increased irradiation of the secondary.  More recently \cite{Hameury2017} performed numerical simulations of truncated accretion discs  around magnetised white dwarfs. They conclude that 
 the more infrequent and short outbursts observed in long-period IPs (compared to dwarf nova) cannot be attributed to the thermal-viscous instability of the accretion disc, but instead have to be triggered by an enhanced mass-transfer from the secondary, or, more likely, by some instability coupling the white dwarf magnetic field with that generated by the magneto-rotational instability (MRI) operating in the accretion disc. 
 
\subsection{The 'circularly polarized' Nova remnants}

\cite{stockman1988} reported the detection of circular polarization in V1500 Cygni, the remnant of nova Cygni 1975, and suggested  that the pre-nova system was a AM Herculis-type magnetic variable that became asynchoronous as a result of the nova explosion. V1500 Cygni has since synchronized \citep{harrison2016}. 

\cite{rodriguez2003} reported the detection of circular polarization in RR Cha, the remnant of Nova Chamaeleontis 1953. Its circular polarization appears not only to be modulated  on the WD spin period but also on the harmonics of the positive super hump  period. RR Cha is an eclipsing system  but little work has been done on this object, probably because of its faintness ($v>18$) and extreme southern declination ($\delta \simeq -82^{\rm o}$).

\subsection{Magnetic field strength}
As pointed out by \cite{Bernardini2017}, if the system is currently in spin equilibrium, as is thought from the inferred mass ratio \citep{Bernardini2017} and the value for $P_{spin}/P_{orb}$ = 0.04, then the magnetic moment of the white dwarf is estimated to be $\sim$5 $\times$ 10$^{33}$ G cm$^3$. Assuming a minimum white dwarf mass $\geq$ 1 $M_{\sun}$ (estimated from post shock region and MEKAL code modeling \citep{Bernardini2017}) leads to a magnetic field estimate of $\geq$ 15 MG, consistent with it being a magnetic CV. Once sychronised, the object will become a polar.   

\section{Summary and Conclusions}

We have detected spin modulated circular polarization from IGR J17014$-$4306, which is the  remnant of Nova Scorpii  1437 A.D., with a peak-to-peak variation of between zero and negative $\sim$3-4\%.  Based on the polarimetry, we propose that it is a disc-fed intermediate polar. The detection of negative circular polarization only is  consistent with the lower accreting magnetic pole being obscured by the accretion disc which results in a net polarization.

IGR J17014$-$4306/Nova Sco 1437 A.D. is a significant object for further studies. Its long orbital period, high-inclination and eclipsing geometry not only helps to constrain the binary parameters but presents opportunities for further detailed investigations using techniques such as Doppler tomography \citep{marsh1982, steeghs2003, kotze2016}, eclipse mapping of the stream, curtains and disc e.g. \citep{hakala2002} and Stokes imaging \citep{potter2004}.

Longer term monitoring of the spin period, to determine the spin equilibrium state of the system, will be important. Combined with an estimate of the magnetic moment of the white dwarf, this  will  help to establish its evolution. Simulations by \cite{norton2004} and \cite{norton2008} have shown that high magnetic moment IPs with long orbital periods will evolve into polars. Low magnetic moment IPs with long orbital periods however have 3 possible  scenarios; (1) they evolve into EX Hya type systems at short orbital periods below the period gap with $P_{spin}/P_{orb}$ > 0.1. These systems may have low magnetic field strength secondaries and so would avoid synchronization (2) they evolve into low field strength polars which are possibly unobservable or (3)  evolve into conventional polars: their 'true' magnetic field strength is  buried by a high accretion rate which re-surfaces when mass accretion turns off. Detailed multi-filtered or spectro-polarimetric observations will help determine the magnetic field strength and hence  differentiate between these possible scenarios.

Long term monitoring will also  reveal the frequency and nature of any outbursts.

\section*{Acknowledgements}

This material is based upon work by the authors which is supported financially by the National Research Foundation (NRF) of South Africa. 









\bsp	
\label{lastpage}
\end{document}